\documentclass[aps,prl,floatfix]{revtex4-1}
\usepackage{graphicx}

\begin{document}

\title{Extremely high-aspect-ratio ultrafast Bessel beam generation and stealth dicing of multi-millimeter thick glass} 

\author{R. Meyer, L. Froehly, R. Giust, J. Del Hoyo, L. Furfaro, C. Billet and F. Courvoisier}
\email{francois.courvoisier@femto-st.fr}
\affiliation{FEMTO-ST institute, Univ. Bourgogne Franche-Comt\'e, CNRS,
15B avenue des Montboucons, 25030 Besan\c{c}on, cedex, France.}


\begin{abstract}
We report on the development of an ultrafast beam shaper capable of generating Bessel beams of high cone angle that maintain a high-intensity hot spot with subwavelength diameter over a propagation distance in excess of 8~mm. This generates a high-intensity focal region with extremely high aspect ratio exceeding 10~000:1. The absence of intermediate focusing in the shaper allows for shaping very high energies, up to Joule levels. We demonstrate proof of principle application of the Bessel beam shaper for stealth dicing of thick glass, up to 1~cm. We expect this high energy Bessel beam shaper will have applications in several areas of high intensity laser physics.
\end{abstract}


\maketitle 

Glass and transparent dielectrics are ubiquitous in modern technology. They are used for consumer electronics, microelectronics, automotive, construction. High speed and high quality cutting of thin and thick glass is therefore an important technological problem.
Interestingly, the recent development of stealth dicing of glass has enable to cleave glass at speeds in the range of 10 to 100~cm/s using lasers with a high repetition rate of several 100's kHz \cite{Mishchik2017, Dudutis2018a}.
Stealth dicing is a two-step technique where the first step consists in generating with individual ultrafast laser pulses a series of high aspect ratio nanochannels which define a weakening plane, serving as a fracture initiator. The second step consists in stressing the material, for instance with a small bending, which is generally sufficient to cleave the glass along the pre-defined plane. The process is ablation-free, does not generate debris and is extremely fast.

Infrared ultrafast Bessel beams are ideal tools to process transparent materials with high aspect ratio such as index modifications \cite{Ye2013,Mikutis2013} or high aspect ratio nanochannels and voids \cite{Bhuyan2010,Rapp2016a}. They are formed by the cylindrically-symmetric interference of plane waves with wavevectors distributed on the generatrix of a cone \cite{Durnin1987}. In the nonlinear regime at high intensities, Bessel beams are quasi distortion-free, provided the cone angle is sufficiently high \cite{Polesana2008}. The high stability of the Bessel beams and the confinement of the intense laser-matter interaction makes it possible to create with a single laser pulse a nanochannel with a diameter typically ranging between 200 and 800~nm \cite{Bhuyan2010} and has led to a number of advances in terms of materials processing via bulk excitation \cite{Courvoisier2016,Stoian2018}.
Stealth dicing using Bessel beams or filaments is now widely used for glass and sapphire separation \cite{Ahmed2013,Bhuyan2015,Mishchik2016_2, Rapp2017}. Until here, stealth dicing has been limited to thicknesses of typically sub-mm. The limit is the available Bessel beam length.

Bessel beams have also a number of different other applications in the field of nonlinear optics \cite{Klewitz1998, Dubietis2007} where they are expected to provide natural tools for amplification with wideband tunability via Kerr instability \cite{Nesrallah2018}, high intensity laser physics \cite{VanDao2009}. Ultrafast Bessel pulses are also emerging in the field of particle acceleration \cite{Hafizi1997, Kumar2017} because the interference creates along the optical axis a high intensity peak that velocity can be tuned and that can even exceed speed of light \cite{Alexeev2002,Zamboni-Rached2003,Clerici2008,Bowlan2009,Froehly2014, Turnbull2018}. Therefore, Bessel beam shapers that can sustain high intensities and high energy over several millimeters are desirable.

\begin{figure}[htb]
    \centering
    \includegraphics[width=0.5\linewidth]{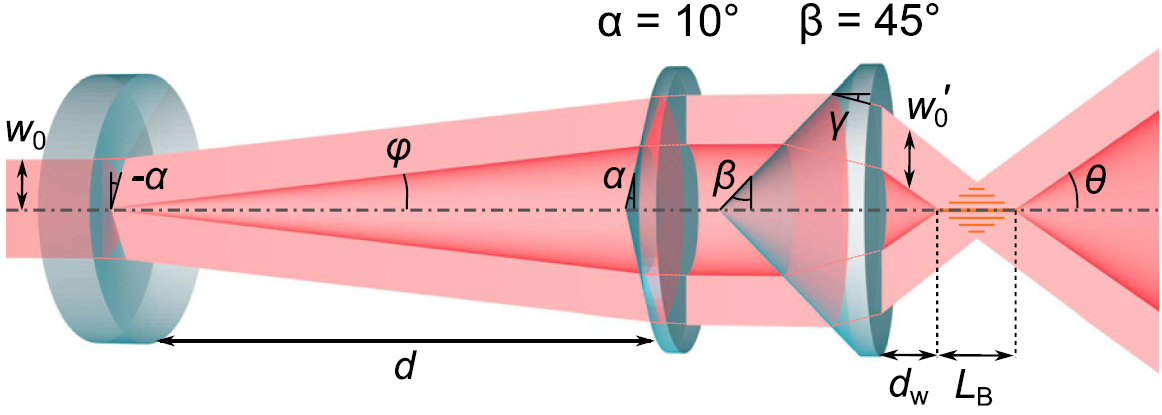}
    \caption{Schematic representation of the beam shaper which is formed by a pair of $\pm10^{\circ}$ axicons and a third axicon with $45^\circ$ wedge angle. The Bessel beam of length $L_\mathrm{B}$ and angle $\theta$ is generated at a working distance $d_\mathrm{w}$ from the last axicon.}
    \label{fig:concept}
\end{figure}

At present, most of Bessel beam shaping techniques for high intensity applications are based on the imaging of an initial Bessel beam formed via an axicon, a spatial light modulator or a diffractive optical element, which is then de-magnified using relay lenses \cite{Bhuyan2010a,Froehly2014,Mitra2015,Mishchik2017}. This approach has several benefits: the Bessel beam is created at a distance from the last optics, enabling a working distance to process thick materials or to realize the generation of the Bessel beam with a smooth injection in the nonlinear medium \cite{Polesana2007}. In addition, the de-magnification factor of the imaging system effectively increases the cone angle of the Bessel beam. This makes possible generating highly focused beams even if the initial shaping element has relatively low spatial frequencies or even if it is difficult to fabricate high angle axicons \cite{Boucher2018}.

However, this approach has important drawbacks. First, the Bessel beam length after imaging is reduced by the square of the magnification factor, which drastically reduces the length that can be reached when high angles are needed and high magnifications used. Second, and most importantly, the imaging techniques make the first Bessel beam prone to distortions during its propagation in air and increase the risks of laser damage. Indeed, during imaging, the Bessel beam field is Fourier-transformed several times. The Fourier transform of a Bessel beam is an annulus. Its width is proportional to $1/L_\mathrm{B}$, $L_\mathrm{B}$ being the length of the Bessel zone \cite{Jarutis2000}. Therefore, high intensities can be reached either within the initial Bessel beam, and/or in the relay optics. Kerr effect, thermal lensing, or optical damage occur when high peak power and high average power pulses are used.

Here, we develop a Bessel beam shaper that has no intermediate focus and which is capable of handling extremely high energies. It improves the Bessel beam zone length by two orders of magnitude for a high angle of $23^{\circ}$ similar to state of the art for single shot nanochannel machining \cite{Bhuyan2010}. Its working distance is adjustable, and the full system is much more compact than those involving relay lenses. We experimentally characterize the Bessel beam distribution up to 1~mJ in air and show it is constant. We demonstrate a proof of principle application to stealth dicing of thick glass, where we reached cleaving up to 1~cm thick soda-lime glass.

The laser source is a Ti:Sapphire Chirped-Pulse Amplified (CPA), Coherent Legend USP, emitting $\sim$~30~fs pulses at a central wavelength of 800~nm, pulse energy of 5~mJ and repetition rate 1~kHz. The pulses can be temporally compressed up to the Fourier-transform limit or stretched using the compressor of the CPA and were characterized just before the beam shaper using an autocorrelator. The concept of the Bessel beam shaper is shown in Figure \ref{fig:concept}. We use a combination of three high-purity fused silica axicons: the first two are respectively negative and positive with the same wedge angle $\alpha$. This transforms the input Gaussian beam, with waist $w_\mathrm{0}=4$~mm ({\it i.e.} radius at $1/e^2$) from the laser source into a thick annulus of collimated light propagating parallel to the optical axis. In the framework of geometrical optics, the width of the annulus is the waist $w_\mathrm{0}$ of the input Gaussian beam; the diameter is determined by the axicons wedge angle and the distance $d$ between the first two axicons. The cone angle $\theta$ is determined only by the wedge angle $\beta$ and index $n_\mathrm{ax} = 1.45$ of the last axicon:
\begin{equation}
\theta = \arcsin \left( n_{\mathrm{ax}}\sin \left( \beta - \arcsin \left( \frac{\sin \beta}{ n_{\mathrm{ax}}} \right) \right) \right)
\end{equation}

The typical length of the Bessel zone $L_\mathrm{B}$ is determined by:
\begin{equation}
L_\mathrm{B} = w_\mathrm{0} (1 + \tan\beta \tan \gamma)/\tan\theta
\end{equation}

Similarly, the working distance $d_ {\mathrm{w}}$ is evaluated from geometrical optics by:
 \begin{equation}
     d_\mathrm{w} = \frac{d \tan \varphi \left(1 + \tan \beta \tan \gamma \right) - e_\mathrm{ax} \tan \gamma }{\tan \theta}
 \end{equation}
in which $e_\mathrm{ax}$ stands for the third axicon thickness (tip to plane).

\begin{figure}[tb]
   \centering
    \includegraphics[width=0.5\linewidth]{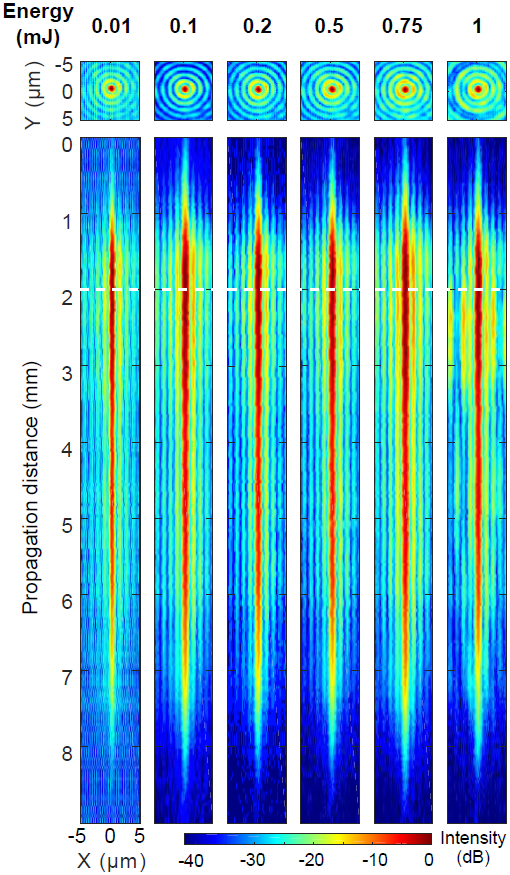}
    \caption{Fluence distribution maps of the Bessel beam in air for different energies from 12~$\mathrm{\mu J}$ to 1~mJ. The maps are displayed in logarithmic scale. The white dashed line on longitudinal sections shows the plane where the maximum intensity is reached, for which we show the beam transverse profiles.}
    \label{fig:beam_sections}
\end{figure}

With the experimental values $\alpha=10^{\circ}$, $\beta$~=~45$^{\circ}$, $e_\mathrm{ax}$~=~17.8~mm and $d$~=~10~cm, we get $\varphi$~=~4.6$^{\circ}$, $\gamma$~=~15.8$^{\circ}$, $\theta$~=~23.3$^{\circ}$, $d_\mathrm{w}$~=~12.2~mm and $L_\mathrm{B}$ ~=~9.7~mm. Our concept is extremely compact because the full length of the beam shaper is $\sim$15~cm, which is much smaller than the Bessel beam shapers based on relay imaging, which length typically exceeds 1~m \cite{Froehly2014}. Because of the high angle of the last axicon, it is oriented with the tip to the laser source to prevent total internal reflection. In this configuration, the distance $d$ must be sufficiently large so that the Bessel beam is formed out of the axicon.

We note that a close concept has been developed by another group \cite{Chebbi2010}, used in Optical Coherence Tomography (OCT) imaging \cite{Weber2012} and recently applied to induce up to 10~mm long modifications in glass \cite{Bergner2018}. However, the latter concept involves a first axicon in focusing geometry such that non-linearities, thermal lensing and optical damage, that we aim to avoid here, might happen for high average input power. The detrimental disruptions in the modifications that are reported to prevent cleaving, might also arise from a too low cone angle and/or detrimental nonlinearities \cite{Polesana2008}.

 We remark that all along the optical path in the beam shaper, the pulse energy is spread over areas that remain on the order of a few cm$^2$, such that with a typical damage threshold of optics of several J/cm$^2$, the beam shaper is expected to handle extremely high pulse energies close to Joule level. With 1~J illumination at 50~fs pulse duration, the peak intensity reached in the Bessel beam would be on the order of $10^{18}$ W.cm$^{-2}$, which is relevant for high-energy physics applications. Similarly, because the pulse energy is quasi uniformly spread over the axicon' surfaces, thermal lensing is largely reduced and its potential focal length would be large, with negligible impact on the Bessel beam structure.

 Experimental characterization of the ultrafast beam was performed via an imaging setup made of two lenses ($f_\mathrm{1}$ = 3.6~cm, $f_\mathrm{2}$ = 1~m) in confocal configuration such that the beam is imaged onto a camera. The magnification of this imaging setup is 27.4. The first lens is 2~inches diameter so that the imaging has a high numerical aperture of 0.58, exceeding the Bessel cone angle. The longitudinal position of the imaging setup is controlled by a motorized translation stage. This allows for scanning the beam over a range exceeding 2~cm. Neutral density filters are placed in the optical path of the imaging setup so as to avoid saturation of the CCD sensor. The damage threshold of the imaging setup is limited to an output pulse energy of 1~mJ (because of the intermediate focusing involved in the imaging), but we could operate the Bessel beam shaper up to the maximal pulse energy available, {\it i.e.} 5~mJ input pulse energy.

Figure \ref{fig:beam_sections} shows experimental characterizations of the beam with different input pulse energies ranging between 12~$\mathrm{\mu}$J and 1~mJ with compressed 50~fs pulses. The characterization is shown with a log scale so as to enhance the visualization of the low-intensity parts and show the high quality of the beam even outside the central lobe. We see the high parallelism and roundness of the profiles in the cross-cuts. The quality of the axicon manufacturing is an important parameter. Imperfect axicons generate a non-diffracting intensity pattern with multiple hotspots unusable for applications. The Bessel beam has a homogeneous transverse distribution over its $\sim$ 8~mm range in air, in agreement with the model described above. This makes the aspect ratio of the beam to be of $>$10~000:1 because the central spot diameter is 740~nm FWHM. The aspect ratio is two orders of magnitude higher than previously achieved with telescopic arrangements for the same cone angle of $23.3^{\circ}$. 

The beam profile does not vary when the pulse energy is increased, evidencing the absence of distortion in the optical system. In addition, we have experimentally verified that varying $d$ and $w_\mathrm{0}$ respectively changes the working distance $d_\mathrm{w}$ and the length of the Bessel zone $L_\mathrm{B}$, without modifying the beam transverse cross-section.

\begin{figure}[t]
    \centering
    \includegraphics[width=0.5\columnwidth]{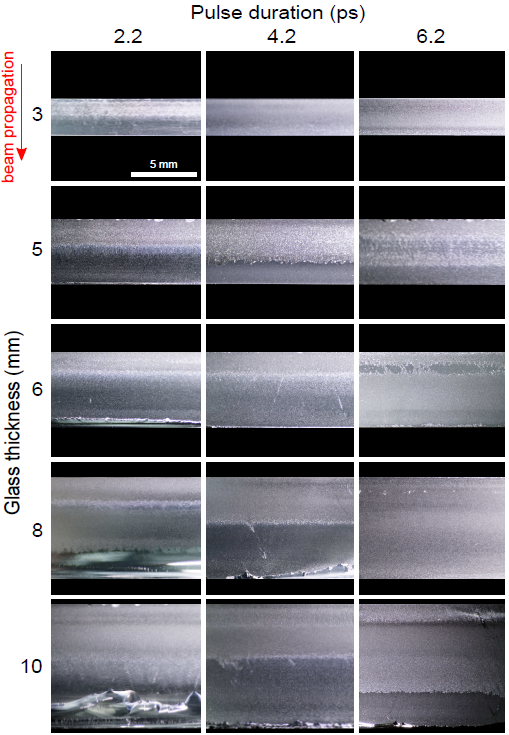}
    \caption{Images of the cleaved edges of glass for thicknesses varying between 3 to 10~mm and for laser pulse durations of 2.2, 4.2 and 6.2~ps. Chipping occurs at the rear surface of the samples for shorter pulse durations and tends to disappear for the longer pulse durations.}
    \label{fig:cleaved_edges}
\end{figure}

Now, we demonstrate a proof-of-principle application of the high energy Bessel beam shaper to stealth dicing of thick glass. For stealth dicing, high cone angles have been shown essential for high quality cleaving \cite{Mishchik2017}, but one of the key limitations for thick glass dicing is to generate material modification on a significant part of the thickness of the material. We demonstrate here that the multi-mm long Bessel beam allows for glass separation up to 10~mm soda-lime glass.

The parameter space is potentially very large and we restricted our study by choosing a fixed translation speed of 5~mm.s$^{-1}$ with 1~kHz repetition rate, so that individual Bessel pulses create material modifications separated by $5~\mathrm{\mu}$m, as found nearly optimal in other studies \cite{Bhuyan2015, Meyer2017}. We used a fixed input pulse energy of 2.5~mJ and varied the input pulse duration in the picosecond regime since pulse duration was evaluated as an important parameter for material excitation. Specifically, material modifications induced by 50~fs pulses were uncleavable. In contrast, picosecond durations have been reported to enhance cleavability in stealth dicing \cite{Lamperti2018, Bhuyan2015}.

We processed $10\times10$~cm$^2$ soda-lime samples of thickness varying between 3 and 10~mm for 3 different pulse durations. Our procedure for stealth dicing was the following. The samples were first laser-processed in single pass. The onset of the Bessel beam was positioned at $\sim$ 0.5~mm before the sample front surface. Then, the samples were mechanically stressed on a 3-lines bending stage. The experiment was repeated 3 times for each set of parameters. Typical results are shown in Figure \ref{fig:cleaved_edges}. We show macro-photography images of the cleaved edges of the samples.

For the 2.2~ps pulse duration, the 3 and 5~mm thick samples are cleaved without observable chipping. Excellent results are observed to separate glass with thickness up to 10~mm, but for longer pulse durations of 4.2~ps and 6.2~ps. For the shortest pulse duration, chipping is observed for 6 to 10~mm mainly in the vicinity of the rear surface. The chipping area increases with glass thickness. We interpret this result as originating from the decrease of local fluence in the central lobe of the Bessel beam for increasing propagation distance. We note a slight increase in the quality of the results between 4.2 and 6.2~ps for the 8~mm case where we report the absence of chipping on all our samples. For the 10~mm case, chipping extends over a distance of some 100~$\mathrm{\mu}$m along the laser beam direction and the deviation from flatness is typically of the same order of magnitude. We note that this could be improved after investigating a wider set of parameters (position of the beam, pulse energy, pulse duration, etc).  We determined with optical profilometry that the RMS roughness of the cleaved samples for 6.2~ps pulse duration is quasi-constant among the samples, with values in the range [1.00 $-$ 1.25]~$\mathrm{\mu m}$. This is close to the roughness of ground glass (typ. 1.0$\pm$0.2~$\mathrm{\mu m}$).
 We finally note that the translation speed of 5~mm.s$^{-1}$ is relatively low in comparison with state of the art stealth dicing, but we believe that high power lasers with high repetition rate will enable improving this parameter.

In conclusion, we have developed a compact Bessel beam shaper producing high energy pulses shaped with $23^{\circ}$ cone angle, over a propagation distance exceeding 8~mm in air. Experimental demonstration has involved energies up to 5~mJ, but we highlight that this Bessel beam shaper has no intermediate focus such that high average power and pulses with extremely high energies, in the range of several 100~mJ to Joules could be shaped. Using such a beam shaper, we manage to upscale stealth dicing technique up to 10~mm millimeters thick glass. With 6~ps pulse duration, the surface roughness of the cleaved glass is similar to the one of ground glass. Therefore, we expect that this technique can save a lot of the energy used at present to post-processed thick glass after mechanical cleaving. Therefore, we anticipate that our results will impact on applications of thick glass processing as well as on more fundamental research for laser plasma physics and high energy laser physics.

The research leading to these results has received funding from the European Research Council (ERC) under the European Union's Horizon 2020 research and innovation program (grant agreement No 682032-PULSAR), European Union 7th Framework Program under grant agreement 619177 (TiSaTD), R\'egion Franche-Comt\'e and the EIPHI Graduate School (ANR-17-EURE-0002).

\bibliography{bibliography}
\end{document}